# Local track irregularity identification based on multi-sensor time-frequency features of high-speed railway bridge accelerations

Ye Mo[1], Yi Zhuo[2], Shunlong Li[1,*],
[1] School of Transportation Science and Engineering, Harbin Institute of Technology, 73 Huanghe Road, Harbin 150090, Heilongjiang, China
[2] China Railway Design Corporation, Tianjin 300142, China
*Correspondence: lishunlong@hit.edu.cn

**Abstract**

Shortwave track diseases are generally reflected in the form of local track irregularity. Such diseases will greatly impact the train-track-bridge interaction (TTBI) dynamic system, seriously affecting train safety. Therefore, a method is proposed to detect and localize local track irregularities based on multis-sensor time-frequency features of high-speed railway bridge accelerations. Continuous wavelet transform (CWT) is used to analyze the multi-sensor accelerations of railway bridges. Moreover, time-frequency features based on the sum of wavelet coefficients are proposed, considering the influence of the distance from the measurement points to the local irregularity on the recognition accuracy. Then, the multi-domain features are utilized to recognize deteriorated railway locations. A simply-supported high-speed railway bridge traversed by a railway train is adopted as a numerical simulation. Comparative studies are conducted to investigate the influence of vehicle speeds and the location of local track irregularity on the algorithm. Numerical simulation results show that the proposed algorithm can detect and locate local track irregularity accurately and is robust to vehicle speeds.

## 1. Introduction

In the straight section of track lines, the difference in geometric parameters produced by the rail relative to the standard straight track position is called track irregularity[1]. During the operation of high-speed railroads, track irregularities are the primary cause of abnormal train vibration. With the increased train running speed, even a small amplitude of track unevenness may cause strong vibration between wheels and rails. Consequently, track quality assessment has become a hot issue in transportation safety research. It is critical to developing early warning systems that minimize disruptions to the transportation network.

Nowadays, high-speed lines mainly use ballastless tracks, so the rail mainly has shortwave diseases. Moreover, shortwave track diseases are generally manifested as local track irregularities. Garg and Dukkipati[2] describe the various typically local irregularity expressions reported in field measurements. Such irregularity will cause an additional impact force from the wheel-rail force and accelerate the deterioration of the rail. Worse, the additional impact force will be transmitted to the under-rail bridge structure through the bridge-rail interaction point, causing potential damage to the bridge structure. Therefore, shortwave track diseases must be detected and maintained as early as possible.

Regularly detecting the health status of the railway system through the track inspection car is the main traditional method[3-5]. However, this method is inefficient and cannot accurately reflect the dynamic track irregularity under the actual train operation. Researchers found that train vibration responses contain many track conditions and disease information. Kouroussis et al.[6] analyzed the vibration impact on the ground due to vehicle excitation caused by local unevenness on the track. This analysis found that the vibration energy is significantly higher when there is local damage. Therefore, it is of practical significance to study track irregularity based on train vibration response.

Since the measurement of axle-box acceleration (ABA) signals does not require complex instrumentation and the accelerometers can be easily fixed outside the axle box. In recent years, many research teams at home and abroad have identified track operation status by monitoring ABA signals[7-12]. The technology is easy to implement, low cost, and has the potential for real-time monitoring. For example, Li et al.[13, 14] proposed a finite element model-based algorithm for early monitoring of track depressions. They pointed out that ABA is sensitive to short-wave irregularity. Molodova et al.[15] indicated that the peak of ABA and its local frequency characteristics may be quantitatively related to the size of the defect. Lin et al.[8] used EMD and Cohen′s class distribution algorithms to analyze ABA and the proposed algorithm can be applied to track unevenness measurements. Yuan et al.[16] used the CVAE-elliptic envelope method to detect and identify the rail squats. Zhang et al.[17] proposed an algorithm to localize tunnel damage using ABA signals and WPE-CVAE.

The bridge structure health monitoring system (BSHM) is equipped with many sensors at different positions of the bridge structure to monitor the operation status of the bridge[18-20]. Over the past decades, many scholars have conducted in-depth studies on various types of monitoring data in BSHM systems. For example, Bao et al.[21] used sparse time-frequency analysis of cable force acceleration to identify time-varying cable tension. Huang et al.[22] developed a method based on Gibbs sampling to solve the problem of sparse stiffness identification. Li et al.[23] modeled the cable tension ratio model using a Gaussian mixture model to evaluate the state of the diagonal cable.

The BSHM monitoring data contains a wealth of structural state information. When there is a local irregularity in the track structure, the additional local excitation will be transmitted to the bridge structure through the track-bridge interaction point, affecting the train-induced bridge accelerations. Therefore, this study analyzes train-induced bridge accelerations at different measuring points to identify the local track irregularity.

Since the vehicle-track-bridge interaction system is complicated. It is challenging to directly obtain the transfer relationship between the bridge responses and the local track irregularity. This study uses continuous wavelet transform (CWT) to infer the corresponding relationship between the local track irregularity and bridge accelerations at different measuring points. Because CWT has the advantage of resolution in the frequency domain, the sum of the extracted time-frequency features-wavelet coefficients is sensitive to the medium and high-frequency responses caused by local irregularities. Furthermore, it is defined as index 1 to detect local track irregularity. The distances between the measuring points and the local irregularity will affect the sensitivity of the identification index. Therefore, this study comprehensively analyzes the index-1 of multiple measuring points to determine which measuring points are closer to the local irregularity. Then only the measuring points that may be closer to the irregularity are analyzed. The local peak points of index-1 are used as the identification index-2 to locate the possible position of local irregularity.

The remaining paper is organized as follows. Section 2 illustrates the proposed methodology and briefly introduces the excitation of local track irregularity in train-track-bridge interaction (TTBI) system vibration. A numerical verification through TTBI simulation is demonstrated in Section 3. Section 4 draws the main conclusions.

## 2. Theoretical background

### 2.1 Excitation of local track irregularity in TTBI system vibration

With increasing rail capacity and speed, the problem of train-track-bridge (TTBI) interaction has become more prominent. The theory of TTBI dynamics has opened up a new field of railroad system dynamics research, creatively studying relatively independent subsystems as a large unified system[24-27].

The presence of track irregularities changes the wheel-rail contact relationship and impacts the dynamic characteristics of the wheel-rail system. The vehicles vibrate under the excitation of track irregularities and other external excitations. Moreover, the vibration is then transferred to the track and bridge through the wheel-rail contact points, thus forming the dynamic interaction process of the TTBI system[24]. Therefore, track irregularity is considered one of the main sources of self-excited excitation for the TTBI system vibration. Its frequency domain characteristics will significantly affect the dynamic response of Bridges and vehicles. The frequency domain characteristics embody the overall fluctuation state of track irregularity. In this section, the harmonic irregularity curve of a single frequency component is selected as the analysis sample:

$$w(x) = \frac{A}{2}\left(1 - \cos\frac{2\pi x}{l}\right) \quad x \in [a,b] \tag{1}$$

Based on the wheel-track corresponding assumption, the vertical displacement of the wheel can be expressed as:

$$z(t) = y(x,t) + w(x) \tag{2}$$

where, $y(x,t)$ and $w(x)$ are the vertical vibration displacement of the bridge and the displacement caused by track irregularity, respectively.

The vertical force on the simply-supported beam could be simplified as follows:

$$P(x,t) = M_1 g - M_1 \frac{d^2 z(t)}{dt^2} \tag{3}$$

in which, $\frac{d^2 z(t)}{dt^2} \approx \frac{\partial^2 y(x,t)}{\partial t^2} + \frac{\partial^2 w(x)}{\partial t^2}$.

In summary, the additional local force $P_{\cos}(t)$ generated by harmonic irregularity can be expressed as:

$$P_{\cos}(t) = M_1 \frac{\partial^2 w(x)}{\partial t^2} = \frac{2\pi^2 v^2 M_1 A}{l^2} \cos\frac{2\pi vt}{l} = P\sin\bar{\omega}t \quad t \in \left[\frac{a}{v}, \frac{b}{v}\right] \tag{4}$$

where, $P = \frac{2\pi^2 v^2 M_1 A}{l^2}$, $\bar{\omega} = \frac{2\pi vt}{l} + \frac{\pi}{2}$. Eq. (4) shows that the effect of local harmonic irregularity on the beam structure can be equivalent to a moving harmonic load.

From the Duhamel integral, generalized coordinates of the *n*-th mode of the supported beam under the action of $P_{\cos}(t)$ are written as:

$$q_n(t) = \frac{2}{mL\omega_D^n} \int_{a/v}^{b/v} P\sin\bar{\omega}\tau \sin n\omega\tau e^{-\omega_b(t-\tau)} \sin \omega_D^n(t-\tau) d\tau \tag{5}$$

According to the trigonometric transformation formula, Eq. (5) can be further written as:

$$q_n(t) = \frac{P}{mL}\left\{ \frac{(\omega_n^2 - r_2^2) - (\cos r_2 t - A) + 2\omega_b r_2 \sin r_2 t - \frac{\omega_b}{\omega_D^n}(\omega_n^2 + r_2^2)A}{(\omega_n^2 - r_2^2)^2 + 4\omega_b^2 r_2^2} - \right.$$

$$\left. \frac{(\omega_n^2 - r_1^2) - (\cos r_1 t - A) + 2\omega_b r_1 \sin r_2 t - \frac{\omega_b}{\omega_D^n}(\omega_n^2 + r_1^2)A}{(\omega_n^2 - r_2^2)^2 + 4\omega_b^2 r_2^2} \right. \tag{6}$$

where, $r_1 = \bar{\omega} + n\omega$, $r_2 = \bar{\omega} - n\omega$, $\omega = \pi v/L$ and $A = e^{-\omega_b t}\sin\omega_D^n t$.

Thus, the acceleration of the bridge under the action of local harmonic irregularity can be expressed as:

$$\ddot{y}_{\cos}(x,t) = \sum_{n=1}^{\infty} \ddot{q}_n(t)\sin\frac{n\pi x}{L} \tag{7}$$

From Eq. (7), it can be seen that, under the action of moving harmonic force, the mid-span acceleration of the simply-supported beam under the action of moving harmonic force can be regarded as the superposition of three harmonic curves. That is, one frequency is $\omega_D^n$, and the other two frequencies are $r_1$ and $r_2$, respectively. It is shown that when there are local harmonic irregularities in the track structure, the high-speed operation of the train leads to additional medium and high-frequency components in bridge accelerations. These frequency components are related to the train speed and the wavelength of the local harmonic irregularities.

## 2.2 Time-frequency features extraction

Continuous wavelet transform (CWT) performs multi-scale signal refinement by scaling translation operation, which has high resolution and adaptability. For the bridge acceleration signal $x(t)$, its CWT form can be expressed as[28]:

$$W_x(a,t) = x(t) * \psi_a(t) = x(t) * \left(a\frac{d\theta_a(t)}{dt}\right) = a\frac{d}{dt}[x(t) * \theta_a(t)] \tag{8}$$

where, $a$ is the scale parameter, the wavelet function $\psi(t) = \dfrac{d\theta(t)}{dt}$, $\theta(t)$ is a smooth function and satisfies $\int_{-\infty}^{\infty} \theta(t)dt = 1$ and it is a higher order infinitesimal of $\dfrac{1}{1+t^2}$; $\theta_a(t) = a\theta\left(\dfrac{t}{a}\right)$ is the original wavelet function.

The absolute value of the sum of wavelet coefficients on the whole scale can be expressed as:

$$W_x(a,t) = x(t) * \psi_a(t) = x(t) * \left(a\dfrac{d\theta_a(t)}{dt}\right) = a\dfrac{d}{dt}[x(t) * \theta_a(t)] \qquad (9)$$

in which, $n$ is the number of scales.

$$S(b) = \sum_{j=1}^{n} |W_x(a,t)| \qquad (10)$$

The inflection points of $W_x(a,t)$ occur when the signal $x(t)$ changes abruptly. At this time, $S(b)$ has a maximum value. Therefore, the mutation point of the signal can be found when the wavelet coefficient is maximized. From the derivation of section 2.1, it can be seen that local track irregularity is equivalent to the additional harmonic force applied to the simply-supported beam. When the train passes through the local irregularity, the dynamic response of the bridge will have different degrees of mutation here. Therefore, by analyzing the bridge accelerations and then performing CWT, the local irregularity can be detected and located by analyzing the sum of wavelet coefficients $S(b)$.

## 2.3 Local track irregularity identification based on multi-sensor time-frequency features

This study mainly studies the detection and location of local track irregularity. The detection process can be divided into the following steps:

**Step 1:** CWT is performed on the acceleration of multiple measuring points, and the sum of wavelet coefficients Si(b) is extracted as the identification index-1. Among them, index-1 is used to detect whether there is local track irregularity.

**Step 2:** The corresponding threshold Fi is defined for different measuring points based on the baseline condition. When the index-1 of a measuring point i exceeds Fi, it is considered that there is local track irregularity.

**Step 3:** The degree of abrupt variation between multiple measurement points is compared to determine which ones are far from the local irregularity. The data from these measurement points are excluded from the subsequent positioning analysis.

The location of local peaks of index-1 is further extracted for localization, which is recorded as index-2. For trains with multiple cars, their wheels repeatedly pass through local irregularity, which makes index-2 exhibit periodicity in space. The periodicity interval is closely related to the spatial distribution of wheels. When the extracted local peak position conforms to the spatial distribution of the wheel at the spatial interval, the location of the local irregularity can be identified. The flow chart of the algorithm in this paper is shown in Figure 1.

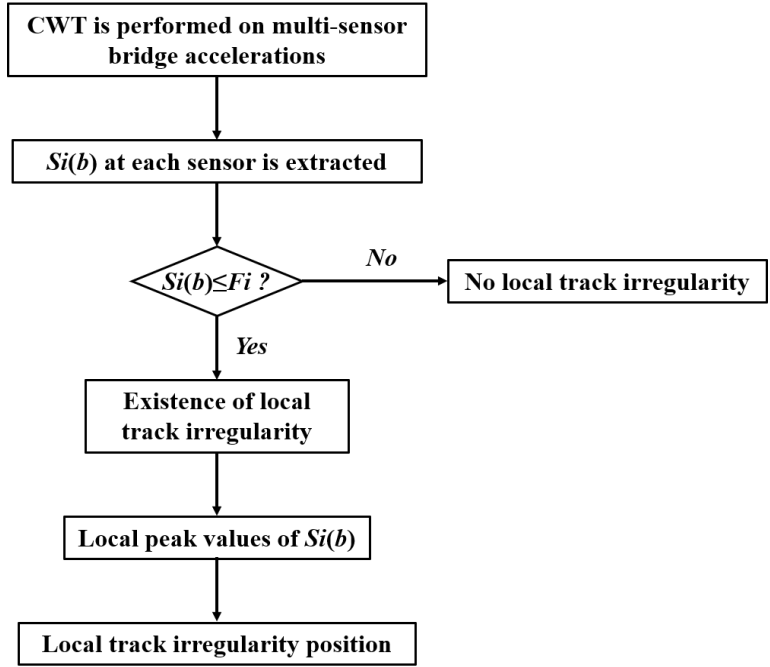

Figure 1. The flow chart of the proposed algorithm

## 3. Numerical verification

In this section, numerical analysis is performed using the Universal Mechanism (U.M.) dynamics model of a high-speed railway simple-supported bridge. The simply-supported beam bridge adopts the single-box single-compartment section. The total length of the bridge is 32.6 m, and its cross-section is shown in Figure 2. The simulation analyzes the vertical acceleration at the five measuring points on the center line of the box girder floor, including the three points where the three sections of the bridge L/4, L/2, and 3L/4 intersect with the center line of the box girder floor, and the two points on the center line of the box girder floor adjacent to both ends of the beam. The sampling frequency $f_s$ is taken as 500Hz.

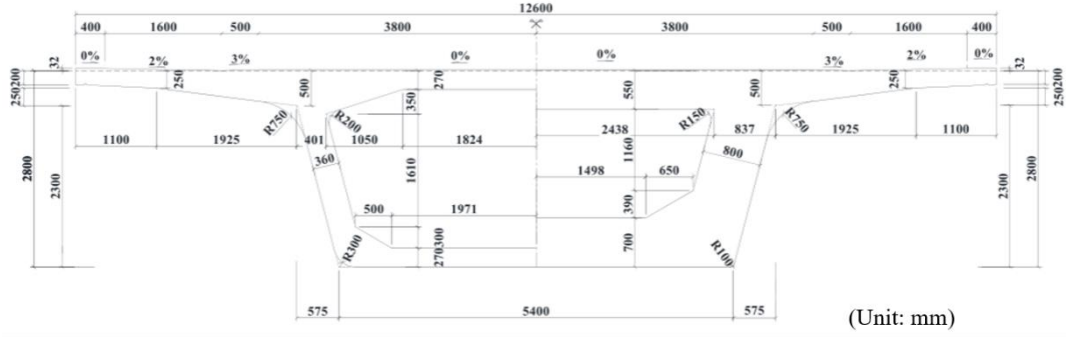

**Figure 2. The cross-section diagram of the box girder**

The vehicle is a three-dimensional space model, using the CRH380A EMU train, which consists of 2 motor cars and 6 trailers. The parameters modeling of the actual train is referenced. The model is modeled concerning the parameters of the actual train.

Moreover, the continuous elastic foundation beam model is selected as the track model in the U.M. program. The rail model can be approximated as a continuous elastic beam. The rail is connected to the under-rail foundation as a parallel connection of a linear spring and damper system while laterally viewed as a series combination. Furthermore, the CN60 rail used for the rails is the Chinese standard 60kg/m rail. Figure.3 shows the dynamic model of the train-track-bridge interaction (TTBI) system.

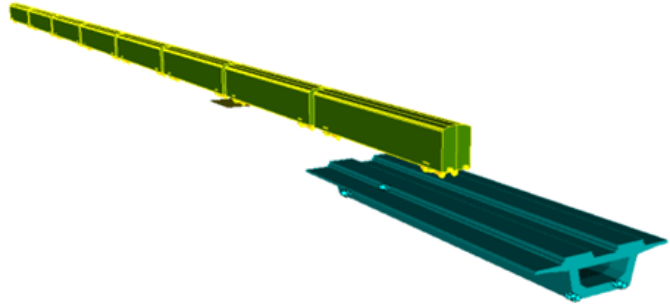

**Figure 3. Multi-body dynamics model diagram of the TTBI system**

### 3.1 Local track irregularity simulation

In this section, the amplitude of local harmonic irregularity is set to 1 mm; the wavelength is 500 mm. Moreover, the local irregularity position is set as 8m, 16m, and 24m, respectively, denoted as single-1, single-2, and single-3 conditions. To further investigate the applicability of the proposed method to multi-point local irregularity identification, multiple-1, multiple-2 and multiple-3 conditions are defined. They are located at 8m + 16m, 8m + 24m and 16m + 24m, respectively. The damaged track irregularity is the sum of initial random and local irregularities, as shown in Eq. (8).

$$w_r(x) = w_i(x) + w_s(x) \qquad (11)$$

in which, $w_r$, $w_i$, $w_s$ are the damaged track irregularity, initial random track irregularity and local irregularity. The CRH2018 track irregularity spectrum is adopted in this simulation as the initialized random track irregularity. Figure 4. exhibits the generated damaged track irregularity of single-1 condition.

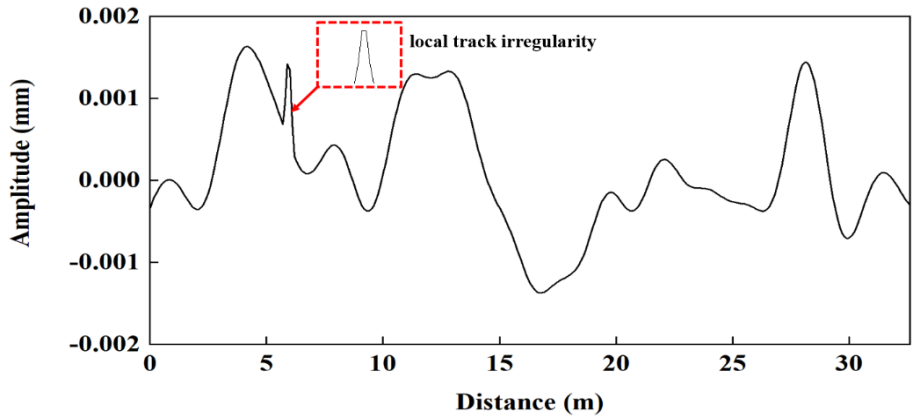

Figure 4. Accumulated track irregularity of single-1

### 3.2 Multi-domain characteristics of bridge acceleration

Figures 5(a)-(e) show the bridge accelerations at the five measurement points. It can be seen that a single local harmonic irregularity causes a sudden change in bridge acceleration in the spatial domain when the vehicle passes through the local harmonic irregularity. Due to the repeated action of multiple trains, the sudden change exhibits an obvious periodicity in the spatial domain. The spatial intervals are related to the spatial distribution of the cars (about the length of a single carriage). The farther the measurement point is from the local unevenness, the less its measured bridge acceleration differs from the base-line model in the spatial domain. Figure.5 suggests that the bridge acceleration response using multiple measurement points can better capture the abrupt changes in acceleration response caused by harmonic irregularities at different locations.

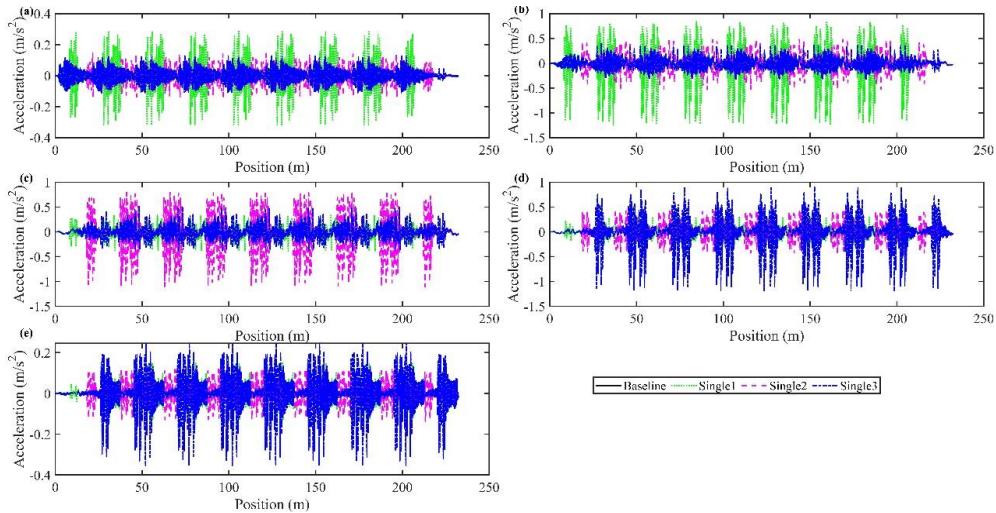

Figure 5. Vertical bridge acceleration in spatial domain

Figures 6(a)-(e) show the frequency domain characteristics of bridge acceleration at different measurement points. It can be concluded that the local harmonic irregularities cause medium and high-frequency vibrations. However, it consists of multiple high dominant frequencies, which indicates that the bridge acceleration caused by local har-monic irregularities is not a single impulse signal. The frequency domain

characteristics of bridge acceleration at different measuring points also differ in sensitivity to local harmonic irregularity. When the measuring point is closer to the local harmonic irregularity, it is easier to capture the influence of local harmonics on bridge acceleration.

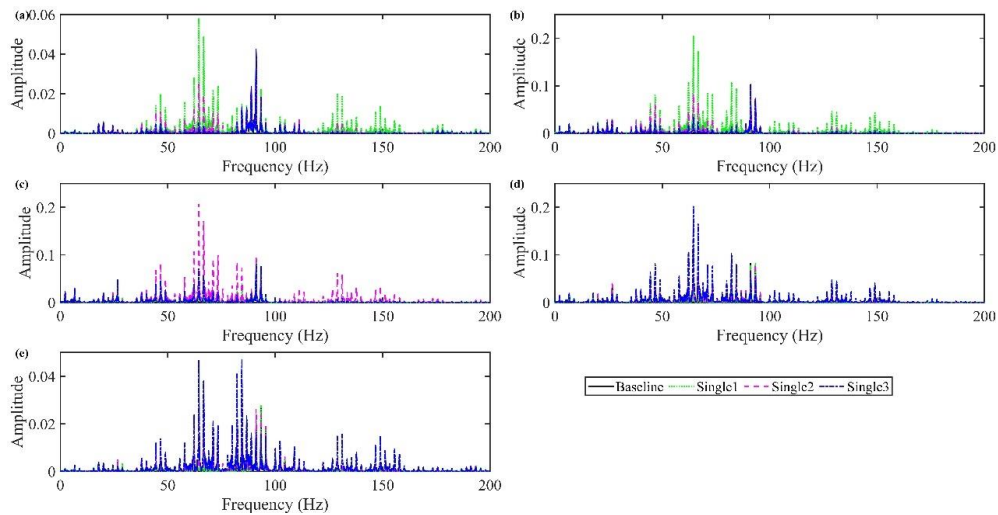

**Figure 6. Vertical bridge acceleration in frequency domain**

Since the frequency domain characteristics lose the time-varying characteristics of the acceleration response, further study of the time-frequency characteristics of multi-sensor bridge acceleration is necessary. Figure 7 is the acceleration wavelet time-frequency diagram at different measuring points. From the diagram, it can be seen that under the excitation of the baseline random track irregularity, there are nine energy concentration areas at measuring points 2-4, which are closely related to the spatial position of the wheelset. Because measuring points 1 and 5 are near the end of the bridge, there are 16 energy con-centration areas. Under the excitation of train load and random track irregularity, the acceleration response of the bridge will produce high-frequency vibration, mainly concentrated at 60-130 Hz.

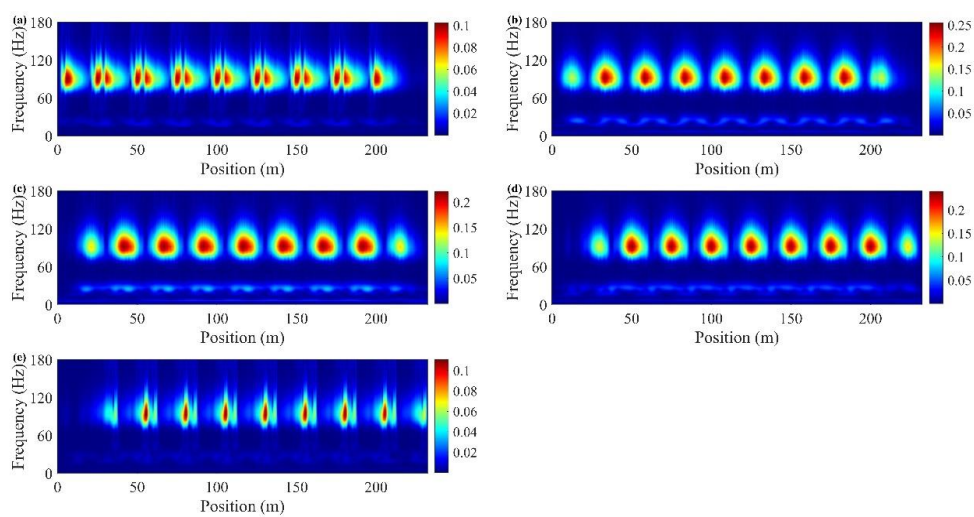

**Figure 7. Time-frequency diagram of bridge acceleration under baseline condition**

Figure 8 is the wavelet time-frequency diagram of bridge acceleration when a local harmonic irregularity is at 8m. It can be implied from Figure.8 that due to local harmonic irregularity, the wavelet time-frequency energy has a significant mutation. The energy mutation generated by measuring points 1 and 2 near 8m is the most obvious. The first position of the mutation is about 8-10 m, and the mutation position is spatially periodic.

In summary, the time-frequency domain can capture the additional response caused by local harmonic irregularity. Therefore, this paper uses the sum of wavelet coefficients at full scale for further analysis.

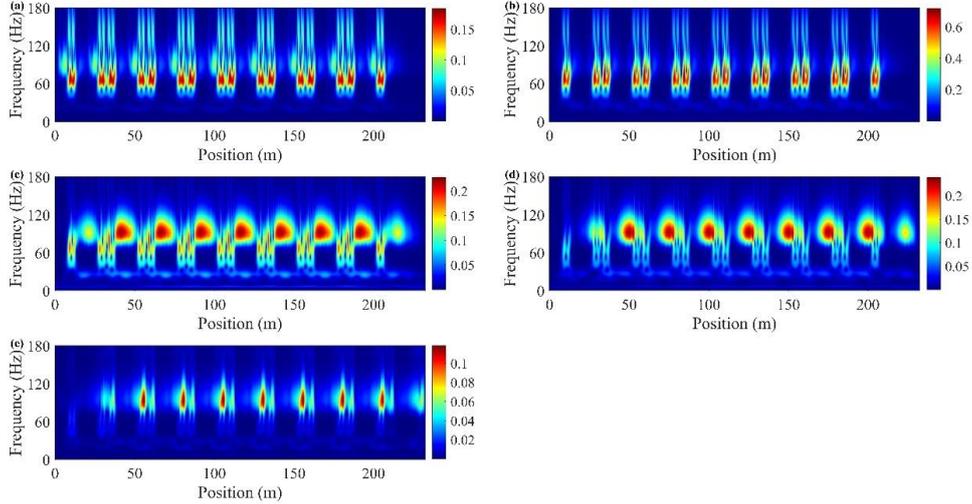

**Figure 8. Time-frequency diagram of bridge acceleration under Single-1**

### 3.3 Data generation and feature extraction

The established TTBI dynamic model is used to generate data for algorithm verification. In order to investigate the effect of vehicle speed on the proposed algorithm, 100 vehicle speed samples are generated within the interval [200,250] km/h and subject to a uniform distribution. It can be summarized in the following steps:

**Step 1:** Track irregularity generation by Eq.(10) and input to the TTBI dynamic model.

**Step 2:** Calculation of vertical bridge acceleration at multiple measuring points.

**Step 3:** Extraction of the sum of wavelet coefficients of the calculated accelerations.

In this section, the acceleration during the train crossing the bridge is chosen for analysis. The different train speeds make the calculated acceleration samples unequal in length. Thus, spatial domain resampling obtains the spatial-domain acceleration with equal spatial intervals.

### 3.4 Local track irregularity detection

In this simulation, threshold $F_i = \mu_i + 3\sigma_i$, $\mu_i$ and $\sigma_i$ are the mean and standard deviation of index-1 extracted from the baseline case, respectively. Figure. 9 shows the sum of wavelet coefficients at full scale for the four conditions at a vehicle speed of 200km/h. The figure shows that the sum of wavelet coefficients of bridge acceleration at each measurement point is much smaller in the baseline condition compared with the local irregularity condition. The sum of wavelet coefficients has a larger peak when the measurement point is closer to the local irregularity. When the measurement point is far from the local irregularity, the peak of the sudden change is smaller, and it is challenging to distinguish whether there is local track irregularity.

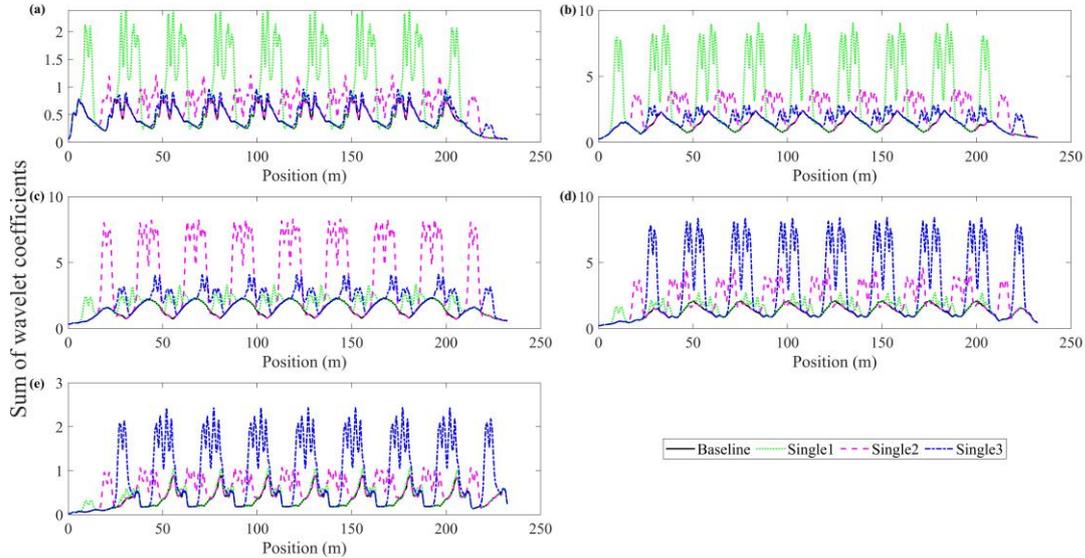

**Figure 9. Sum of wavelet coefficients at vehicle speed 200km/h**

The time-frequency characteristics of bridge acceleration at 250km/h are further analyzed, as shown in Figure 10. Comparing Figure. 9 and Figure. 10 shows that speed has little influence on the results of the proposed index-1. Local track irregularity can be detected by comparing the multi-sensor index-1 values in the baseline condition with other operating conditions.

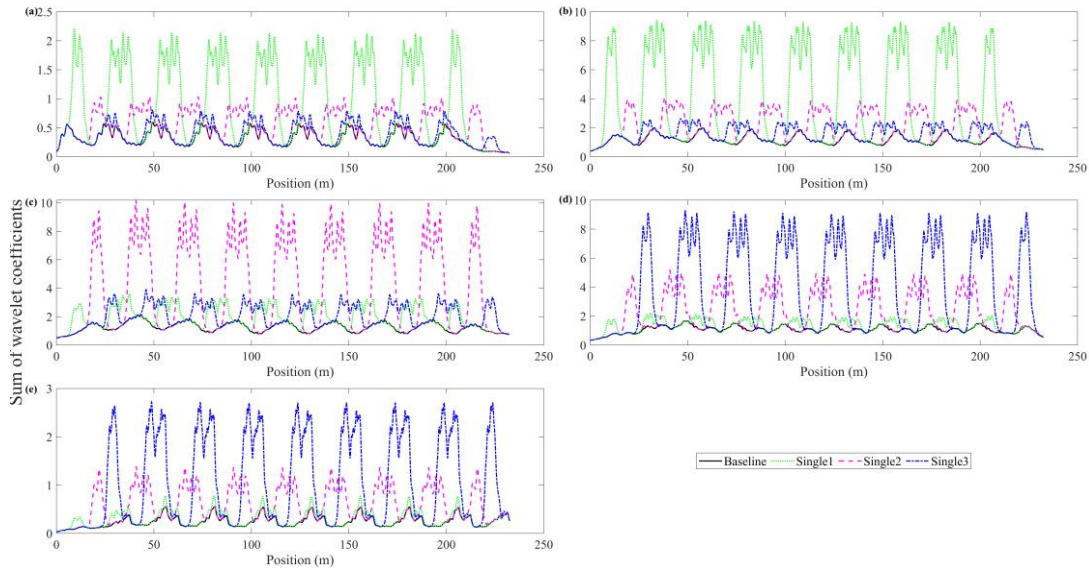

**Figure 10. Sum of wavelet coefficients at vehicle speed 250km/h**

The sum of wavelet coefficients for the multi-point irregularity condition is shown in Figure 11. The wavelet coefficients and the degree of abrupt variation of the multi-point non-smooth condition are significantly larger than those of the single-point non-smooth condition. In addition, the number of local peak points of the wavelet coefficients is also larger.

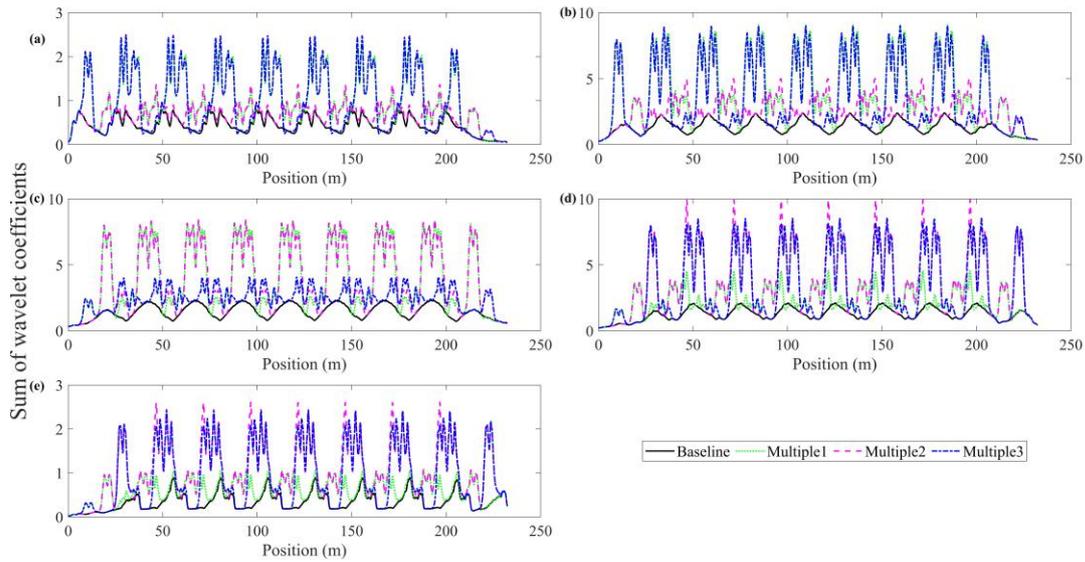

Figure 11. Sum of wavelet coefficients for multi-point uneven conditions at a speed of 200km/h

### 3.5 Implementation cases for local track irregularity localization

The example of two such conditions illustrate the localization process of the algorithm, single-1 and multiple-1, at a vehicle speed of 200km/h. The wavelet coefficients sum curves of Single-1 and Multiple-1 compared with the baseline conditions are shown in Figures. 12-13.

As shown in Figure 12. (a)-(b), under the Single-1 condition, the sum of wavelet coefficients of measuring point 1 and measuring point 2 has the largest mutation degree. It indicates that there is local irregularity near these two measuring points. Similarly, the Multiple-1 condition is then analyzed. From Figure 13., it can be seen that the measuring points 1-3 are relatively close to the local irregularity. In summary, the sum of wavelet coefficients of measurement points 1-2 and 1-3 are selected for locating the local track irregularity of conditions single-1 and multiple-1, respectively.

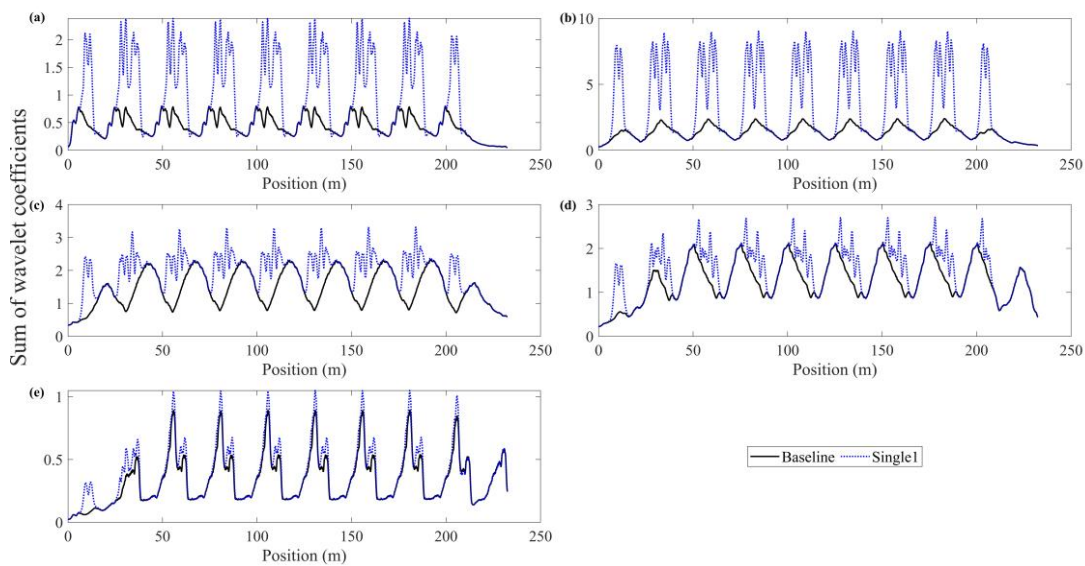

Figure 12. Comparison of sum of wavelet coefficients between Single-1 and the baseline condition

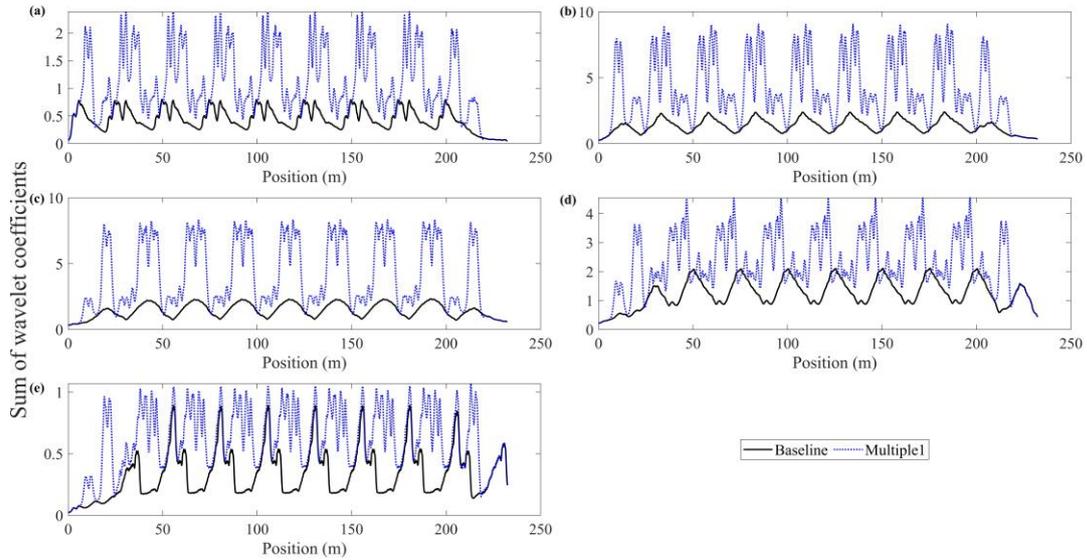

Figure 13. Comparison of sum of wavelet coefficients between Multiple-1 and the baseline condition

The index-2, the local peak of the sum of the wavelet coefficients, is used for positioning. Figures.14-15 are the local peak points identified under Single-1 and Multiple-1 conditions, respectively. The first local peak identified by measuring points 1 and 2 is 9m. The spatial interval of the remaining local peak points is consistent with the spatial distribution of the wheel. It shows that there is a local irregularity near 9m. The identification result is close to the local irregularity position of single point 1.

Similarly, the locations of the peak points shown in Figure 15 are analyzed. From this, it can be seen that measurement points 1 and 2 show a local irregularity near 9m, and analysis of measurement point 3 shows a local irregularity near 19m. The identification results indicate local irregularities at two locations in this case.

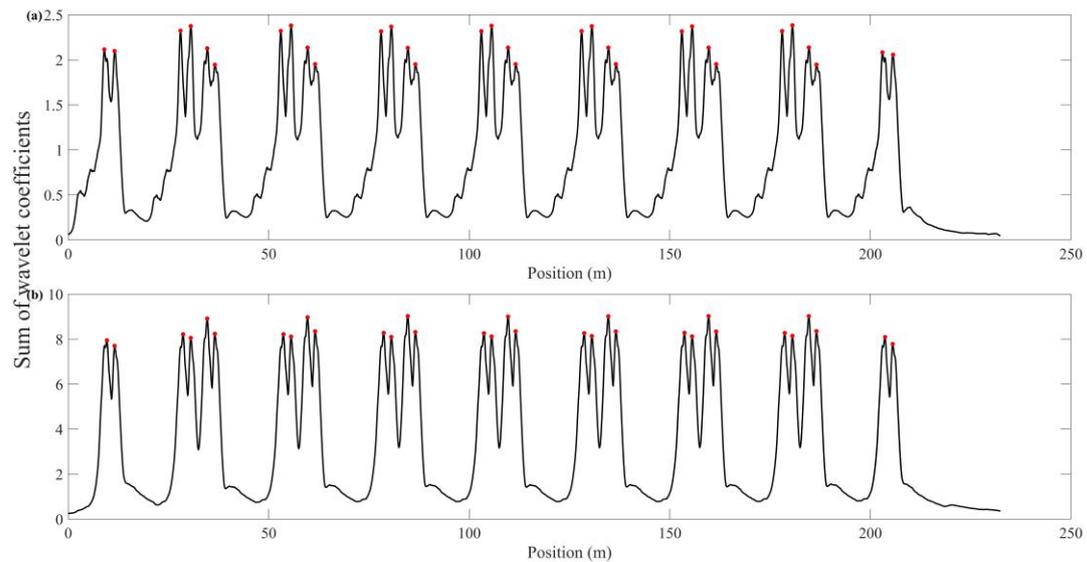

Figure 14. Local peaks of wavelet coefficient sum under Single-1

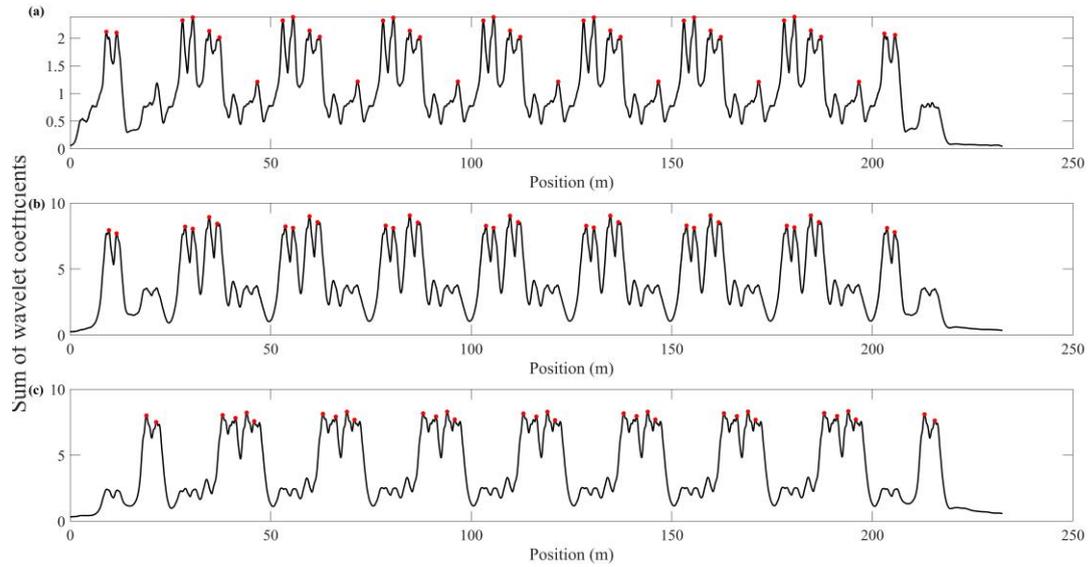

Figure 15. Local peaks of wavelet coefficient sum under Multiple-1

## 4. Conclusion

In this study, a local track irregularity identification method has been developed. The proposed approach utilizes CWT to extract multi-sensor time-frequency features of train-induced bridge accelerations. Since local irregularity will cause sudden changes in the time-frequency energy of bridge accelerations, the extracted features can be used to detect and locate local irregularity. A TTBI dynamic simulation has been used to verify the feasibility and effectiveness of the algorithm. Effects of vehicle speed and location of local irregularity are investigated, through which main conclusions can be drawn as follows:

- The action of the local harmonic irregularity on the bridge structure can be equated to the action of the moving simple harmonic load. Local harmonic irregularities in the track structure will result in additional mid- and high-frequency components in the bridge acceleration response. Moreover, These additional frequency components are related to the train speed and the wavelength of the local harmonic irregularities.
- The sum of the wavelet coefficients in the full scale is used as the local track irregularity detection index-1, which reflects the change of the time-frequency energy of the bridge acceleration with the train running position. When the train passes through the location of local irregularity, the index will have a sudden change in the spatial domain. Moreover, the degree of mutation is related to the distance from the measuring point to local irregularity. When the measurement point is closer to the local unevenness position, the peak of index-1 mutation is greater and vice versa.
- The local peak points of index-1 are used as index-2 to locate local irregularity.

As many carriages pass through local irregularity, the identified local peak points have relatively obvious periodic intervals in the spatial domain, which is related to the position of the wheels in the spatial.
- The two indexes proposed in this paper have relatively strong robustness to train speed and local irregularity position, and index-2 can identify multi-point local irregularity position.

**Author Contribution:** All authors participated in the discussions and made scientific contributions. Conceptualization, Shunlong Li; Formal analysis, Yi Zhuo; Funding acquisition, Shunlong Li; Investigation, Ye Mo; Methodology, Ye Mo; Project administration, Shunlong Li; Software, Yi Zhuo; Supervision, Shunlong Li; Writing – original draft, Ye Mo; Writing – review & editing, Shunlong Li.

**Funding:** The research described in this paper was financially supported by China Railway Design Corporation R&D Program [2020YY340619，2020YY240604], and Fundamental Research Funds for the Central Universities [FRFCU5710051018].